\documentclass[journal,comsoc]{IEEEtran}

\usepackage[T1]{fontenc}
\usepackage{color}
\usepackage{ifthen}
\usepackage{xspace}
\usepackage{paralist}
\usepackage{cite}

\usepackage[pdftex]{graphicx}

\newcommand{\vs}{vs.\ }

\newcommand{\wrt}{w.r.t.\ }

\newcommand{\fakeparagraphnodot}[1]{\vspace{1mm}\noindent\textbf{#1}}
\newcommand{\fakeparagraph}[1]{\fakeparagraphnodot{#1.}}
\newboolean{authnotes}
\setboolean{authnotes}{false}

\ifthenelse{\boolean{authnotes}}
{
\newcommand{\davide}[1]{\footnote{{\bf Davide: #1}}}
\newcommand{\amy}[1]{\footnote{{\bf Amy: #1}}}
\newcommand{\tim}[1]{\footnote{{\bf Tim: #1}}}
\newcommand{\gp}[1]{\footnote{{\bf GP: #1}}}
\newcommand{\elia}[1]{\footnote{{\bf Elia: #1}}}

\usepackage{draftwatermark}
\SetWatermarkLightness{0.85}
}
{
\newcommand{\davide}[1]{}
\newcommand{\amy}[1]{}
\newcommand{\tim}[1]{}
\newcommand{\gp}[1]{}
\newcommand{\elia}[1]{}

}
\usepackage[cmintegrals]{newtxmath}
\usepackage{url}
\begin{document}
\title{Janus: Efficient and Accurate\\Dual-radio Social Contact
  Detection}
\author{Timofei Istomin, Elia Leoni, Davide Molteni, Amy L. Murphy,
  Gian Pietro Picco, Maurizio Griva
  \thanks{T. Istomin, D. Molteni,
    G.P. Picco are with the University of Trento, Italy. E-mail:
    \{timofei.istomin, davide.molteni,
    gianpietro.picco\}@unitn.it.}%
  \thanks{E. Leoni and A.L. Murphy are with the
    Bruno Kessler Foundation, Trento, Italy. E-mail: \{eleoni,
    murphy\}@fbk.eu.}%
  \thanks{M. Griva is with Concept Reply, Turin, Italy. E-mail:
    m.griva@reply.it.}%
}
{}
\maketitle

\begin{abstract}
Determining when two individuals are within close distance is key to
contain a pandemic, e.g., to alert individuals in real-time and trace
their social contacts. Common approaches rely on either Bluetooth Low
Energy (BLE) or ultra-wideband (UWB) radios, that nonetheless strike
opposite tradeoffs for energy efficiency \vs accuracy of distance
estimates.

Janus reconciles these dimensions with a dual-radio protocol enabling
efficient \emph{and} accurate social contact detection. Measurements
show that Janus achieves weeks to months of autonomous operation,
depending on the configuration. Several large-scale campaigns in
real-world contexts confirm its reliability and practical usefulness in
enabling insightful analysis of contact data.

\end{abstract}

\IEEEpeerreviewmaketitle

\section{Social Contact Detection}
\label{sec:intro}

\IEEEPARstart{S}{ocial} distancing is one of the key instruments
available to society for the containment of dangerous viruses like
COVID-19.  In this context, the use of radio-enabled devices, e.g.,
smartphones, or dedicated ``proximity tags'', are advocated by many as
a formidable tool to support \emph{contact detection}: determining
when two individuals come within close distance.

Contact detection supports social distancing along several dimensions,
including:
\begin{inparaenum}[\em i)]
\item \emph{real-time enforcement} of social distancing, e.g.,
  automatically alerting people upon contact detection,
\item \emph{monitoring and recording} of the distance and duration of
  a contact, enabling \emph{offline analysis}, e.g., to trace the
  spread of infection from a diagnosed individual.
\end{inparaenum}

\section{Requirements}

Achieving these goals poses multiple technical challenges. 

An effective contact detection solution should be \emph{reliable}, a
notion with several facets in this context. Obviously, false negatives
(contacts occurring and going undetected) should be
minimized. However, a contact between two individuals is associated
with a \emph{distance} and a \emph{duration}, together discriminating
the danger of contagion; therefore, for detection to be reliable it
must be \emph{accurate} and \emph{timely}. Minimizing false positives
(safe contacts detected as occurring at unsafe distance) is key, as
they may generate unnecessary alarms, undermining user confidence in
the tool, or bias data interpretation and contact tracing. Moreover,
detection must occur within well-defined time bounds, to ensure prompt
user alerting or correctly capture the effective contact duration.

On the other hand, by their nature, these systems must rely on devices
carried by users, thus battery-powered. Therefore, contact detection
must also be \emph{energy-efficient}; an accurate and timely system is
not very useful if its battery depletes in a few hours. The shorter
the lifetime, the higher the maintenance overhead for the user and
therefore the barrier to adoption. This is exacerbated in the common
use case where tags are owned and managed by an organization to ensure
safe working conditions; the cost of frequently recharging hundreds or
even thousands of devices cannot be neglected.

Unfortunately, these system requirements are at odds: an always-on
radio fosters timely detection but also quickly depletes the
battery. Further, they need to be reconciled with specific contact
definitions, which may change depending on country regulations (e.g.,
stipulating different safe distances) or use cases (e.g., factory
floor \vs office buildings). Navigating and optimizing these
constraints demand a system that is easily
\emph{configurable}.

\section{State of the Art and Contribution}

These ideas recently led to a flurry of systems by private companies
and national public entities.

\fakeparagraph{Smartphone-based apps and GAEN} Among these, arguably
the most prominent is the Google-Apple Exposure Notification (GAEN),
an OS-level mechanism exploiting Bluetooth Low Energy (BLE) for
contact detection on mobile phones, at the core of ``COVID apps'' in
many nations. This concerted effort by two major players offers a
formidable penetration in the smartphone user base, but suffers from
several problems.

First, its detection operation is \emph{fixed}: each phone emits a BLE
advertisement every
$\sim$$250ms$ and scans for those from other phones every
4~minutes. This is an understandable compromise, given the inherent
need to provide a one-size-fits-all solution across the entire
installed base and various use cases. Nevertheless, it prevents
alternate configurations striking different tradeoffs between
timeliness and energy consumption.

A more disruptive limitation concerns accuracy. GAEN relies on the
radio signal strength indicator (RSSI) reported upon reception of BLE
advertisements to estimate distance via its relation with signal
attenuation. However, this technique is notoriously affected by
environmental conditions, also contributing to attenuation. This
yields significant estimation errors and ultimately invalidates the
data collected, as recently assessed empirically in real
environments~\cite{leith20:plosone}.

Finally, several privacy concerns have arisen, stimulating a technical
debate about centralized \vs decentralized architectures for sharing
contact data, but also arguably hampering a larger adoption of these
smartphone applications~\cite{ahmed20:survey}.

\fakeparagraph{Proximity tags} Albeit pervasive, smartphones are not
the only (or the best) devices enabling contact detection. Not
everyone owns a smartphone (e.g., many children and elders) and those
who do may be reluctant to give partial access to such an integral
part of their digital life, due to the privacy concerns
above. Finally, the decision whether to participate in contact
detection rests solely with the user, who must explicitly install and
correctly use the corresponding app.

These considerations fueled a market surge of ``proximity tags'',
geared both towards real-time alarms and offline analysis. Unlike
smartphone applications, which target global use, these devices target
situations where the use of tags can be controlled and enforced. For
instance, the Bump~\cite{bump} alerting system recently made the news
as its use was required by all athletes and staff participating in the
London marathon. More menial applications of wider relevance include
monitoring of children (e.g., at school or summer camps) and elders
(e.g., in retirement homes), and ensuring workplace safety.

\fakeparagraphnodot{BLE or UWB?} Several tags on the market are based
on BLE, whose pervasiveness and low energy consumption enable cheap,
long-lasting devices at the price of poor accuracy, as already
outlined for GAEN-based apps. However, once the leap from a smartphone
to a custom tag is made, alternate designs offering better performance
are possible.

This is the case of tags exploiting ultra-wideband (UWB) radios. These
operate on fundamentally different PHY-level principles that enable
distance estimates with an error $<$10~cm, i.e., 1--2 orders of
magnitude less than narrowband radios like WiFi and BLE, significantly
enhancing contact accuracy. UWB localization systems are rapidly
gaining traction and, by yielding accurate and timestamped positions,
indirectly enable contact detection. Nevertheless, they also require
an infrastructure of fixed reference nodes (anchors), implicitly
delimiting the area where detection can occur, with conflicting
tradeoffs of scale \vs effectiveness \vs cost. Therefore, although
hybrid solutions exist~\cite{pozyx,wipelot}, UWB-based proximity tags
typically measure \emph{directly} the distance between two devices via
standard~\cite{std154} or custom-designed~\cite{cao20:6fit}
\emph{ranging} schemes.

Unfortunately, UWB energy consumption is roughly an order of magnitude
higher than BLE, significantly affecting device lifetime. For
instance, the aforementioned Bump system claims only 12~hours of
operation~\cite{bump}; others fail to report lifetime
altogether~\cite{cao20:6fit}.

\fakeparagraph{Janus: A dual-radio approach} 
Named after the god with two faces in Roman mythology, \emph{Janus}
takes the best of BLE and UWB: the low-power consumption of the former and 
the high accuracy of the latter.

BLE retains the central role of discovering devices (i.e., users) in
range germane to the aforementioned approaches. This \emph{continuous}
neighbor discovery is expensive in terms of energy. Imagine an
individual alone in the office for most of the day, obeying social
distancing rules and only seldom interacting with co-workers. Although
there is no one to discover, the channel must be scanned to ensure
timely detection in case a colleague appears. A tag based solely on
UWB would rapidly deplete the battery in this wasteful task, due to
high energy consumption. This does not occur in our dual-radio
architecture, where continuous neighbor discovery is performed by the
lower-energy BLE radio, while the higher-energy UWB radio is
\emph{triggered on-demand} solely when contact detection occurs and
distance estimates are required. Moreover, UWB approaches must avoid
collisions among ranging exchanges; for instance, the scheme
in~\cite{cao20:6fit} reports that only 65\% of them are executed
successfully. Janus achieves a near-perfect success rate by
piggybacking on the out-of-band BLE channel information to
\emph{coordinate} UWB ranging exchanges.

This dual-radio approach is largely novel among both research and
commercial devices; only few exist, \wrt which Janus enables
significant advantages. The work in~\cite{biri20:socitrack} exploits
BLE only for neighbor discovery; coordination of ranging exchanges is
performed via UWB in a centralized fashion. This not only yields
significantly higher consumption, but also severely limits the
applicability in the highly dynamic scenarios of practical interest.
Among commercial tags, a few~\cite{ProDongle,Ubudu} use BLE only as an
out-of-band channel to collect data and set
configurations. Wipelot~\cite{wipelot} exploits instead a combination
of UWB and IEEE~802.15.4. However, the latter radio has higher energy
consumption \wrt BLE, enabling smaller savings; a 3-day lifetime is
claimed, in unspecified conditions, while Janus achieves up to 3~weeks
with a short, 2-second detection.

\section{Dual-radio Discovery and Ranging}
\label{sec:janus}

We summarize the two enabling techniques of UWB-based
ranging and BLE-based neighbor discovery, then illustrate how we
exploit them in synergy in our Janus protocol.

\newcommand{\poll}{\textsc{poll}\xspace}
\newcommand{\response}{\textsc{response}\xspace}

\subsection{Building Blocks}

\fakeparagraph{UWB: Ranging} Ultra-wideband has returned to the
forefront of research and market interest after a decade of oblivion,
thanks to small, cheap, and energy-savvy new UWB impulse radio chips,
spearheaded by the DecaWave DW1000 we use here. The use of very short
pulses ($\leq 2ns$) reduces power spectral density and interference
from other radios while improving propagation through obstacles. The
large bandwidth yields superior ranging accuracy via excellent time
(hence distance) resolution, enabling receivers to precisely timestamp
the signal time of arrival and discriminate it from multipath.

Two-way ranging (TWR) is commonly used to estimate distance between
two UWB nodes. The simplest variant, single-sided TWR (SS-TWR) is part
of the IEEE~802.15.4 standard~\cite{std154} and requires a 2-packet
exchange between an initiator and a responder.  The initiator
transmits a \poll packet to the responder, which replies with a
\response after a known delay. This packet includes the timestamps
marking the reception of \poll and transmission of \response that,
along with the TX/RX timestamps at the initiator, enable it to compute
the time-of-flight and estimate distance multiplying by the speed of
light in air.

\begin{figure*}[!t]
  \centering
  \includegraphics[scale=0.8]{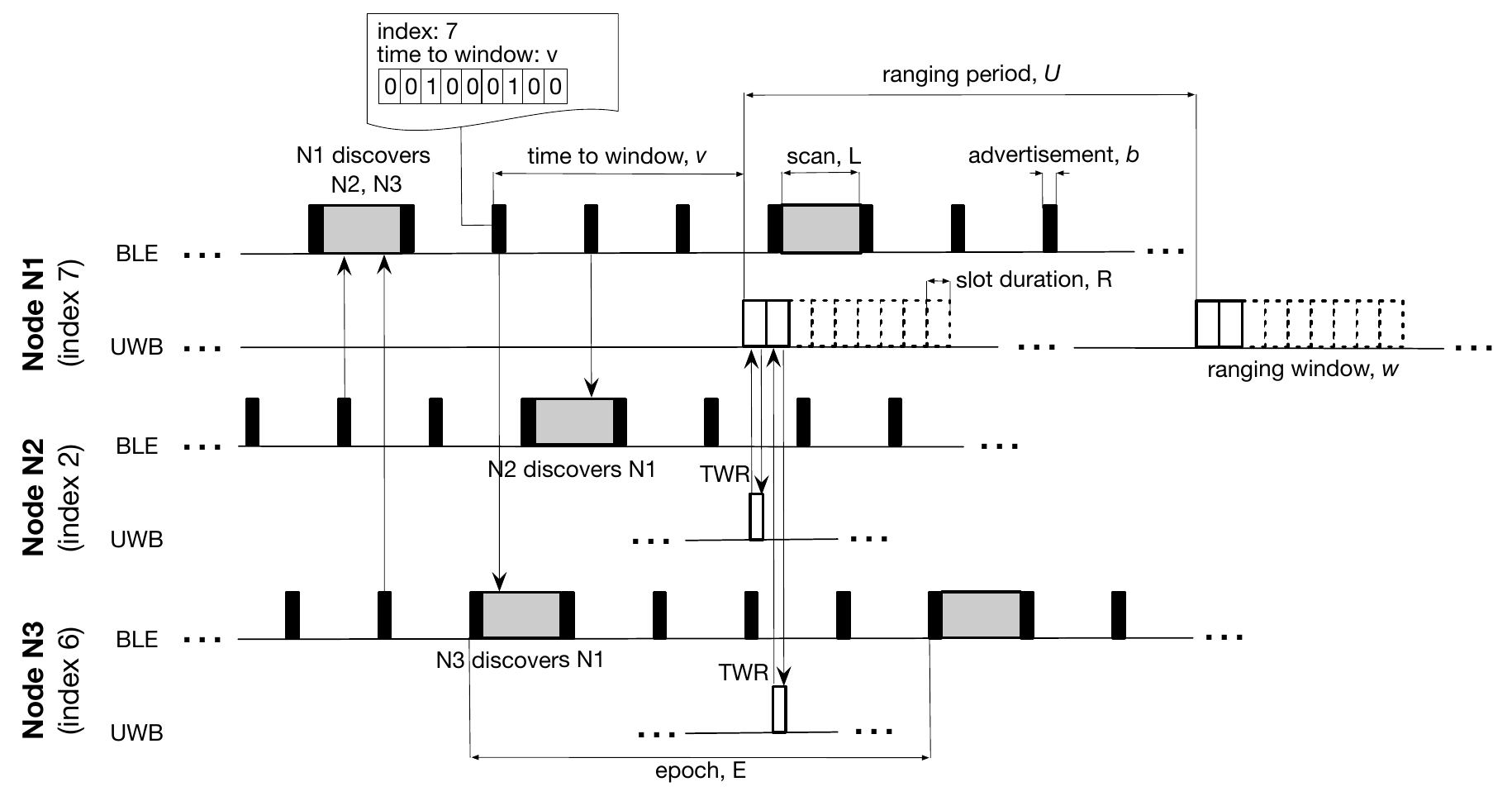}
  \caption{The Janus protocol. The illustration is a complete account
    only for node $N1$. Although $N2$ and $N3$ also discover each
    other during their scans and schedule their own ranging windows,
    the corresponding portions of the schedule are omitted here for
    readability. }
  \label{fig:janusProtocol}
\end{figure*}

\fakeparagraph{BLE: Neighbor discovery} We exploit
BLEnd~\cite{julien17:blend}, a state-of-the-art neighbor discovery
protocol developed in our group. In any BLE-based discovery protocol,
a node must transmit advertisements to announce itself and scan
(listen) for those from other nodes. BLEnd provides the schedules for
these two operations, taking into consideration the expected density
of the neighborhood, as this affects the collisions among
advertisements, leading to missed contacts. The result is an
energy-efficient protocol with well-defined discovery guarantees.
A companion optimizer tool simplifies the task of configuring protocol
parameters towards a given detection latency or energy target while
achieving predictable performance.

\subsection{Exploiting BLE and UWB in Synergy}

Janus merges BLE-based neighbor discovery and UWB ranging into a single
energy-efficient protocol (Figure~\ref{fig:janusProtocol}).

BLE scans and advertisements are executed periodically according to
the BLEnd protocol. The discovery period $E$ (epoch, in BLEnd) and the
scan duration $L$ (which also determines the interval between two
advertisements) are determined by the BLEnd optimizer to meet a
discovery latency while maximizing lifetime; the advertisement
duration $b$ is instead fixed by the BLE radio.

UWB ranging also repeats periodically. Once a node has discovered at
least one neighbor, it schedules its own ranging window with a period
$U$, randomized by a small jitter to avoid long-lasting overlaps with
those of other nodes. Each window contains one slot of size $R$ per
discovered neighbor, resulting in a \emph{dynamic} window duration.
To inform neighbors when to initiate ranging, a node adds in the
payload of BLEnd advertisements:
\begin{itemize}
\item its node index, unique in the neighborhood;
\item the time $v$ to the beginning of the next ranging window,
  updated for each advertisement;
\item a bitmap indicating the slot allocation for ranging.
\end{itemize}
When this information arrives at a node $N2$ in the BLE advertisement
from $N1$, $N2$ performs ranging in its slot allocated in $N1$'s
window,
obtaining the distance between the two nodes. Thanks to
the bidirectional discovery enabled by BLEnd, the dual process occurs at $N1$
and for all neighbors (not shown in Figure~\ref{fig:janusProtocol}).

Slots are allocated for neighbors at the end of each ranging window
and de-allocated only after a given number of advertisements are no
longer received, indicating the neighbor has moved away. 

\fakeparagraph{Synchronizing with BLE advertisements} Each BLE
advertisement consists of 3~identical packets sent sequentially on
different channels (37$\rightarrow$38$\rightarrow$39).  As each scan
occurs on a single channel, changed after each scan, the scanning node
receives only one of the packets at a fixed time offset depending on
the position in the sequence. However, since the channel sequence is
invariant and the RX channel and inter-packet interval in an
advertisement are known, the node computes the time the first packet
was sent and uses it as reference to schedule ranging.

\fakeparagraph{Node index} The ranging window must schedule a slot for
each neighbor; depending on the deployment, there may be tens of
them. As the schedule must fit into a single BLE advertisement payload
(at most $24B$), identifying nodes by their $6B$ address is
unfeasible. Instead, we identify nodes with a 1-byte index and
advertise bitmaps where a $1$ in position~$x$ denotes a ranging slot
allocated for the node with index~$x$
(Figure~\ref{fig:janusProtocol}).
The slot number is defined as the ordinal number of the 1 in the
bitmap. The figure shows a 9-neighbor schedule, specifying that nodes
with index~2 and~6 are expected to range in the first and second slot,
respectively.

This bitmap must accommodate the maximum expected number of neighbors
and minimize conflict among indexes, discussed next. Therefore, we use
all remaining 104~bits ($13B$) in the advertisement payload.

\fakeparagraph{Resolving node index conflicts} The nodes in a
deployment may be many more than the available node indexes, which
therefore cannot be \emph{globally} unique. Still, indexes must be
\emph{locally} unique, otherwise multiple nodes would share the same
slots and their ranging packets would collide. We developed a conflict
resolution strategy that reassigns indexes upon detecting conflicts.

At bootstrap, nodes select their index randomly. As advertisements
include the sender index, receivers can detect conflicts with their
index; the node with the lower BLE address changes its index randomly,
avoiding those already in use. In case two non-neighboring nodes with
the same index share a neighbor, the latter indicates the conflict in
the advertisement payload, forcing both neighbors to select a
different index.

To help select available indexes, each node caches the bitmaps of all
neighbors; their bit-wise \texttt{OR} with its own schedule yields a
zero for all unused index values.

\section{From a Prototype to a Full-fledged System}
\label{sec:reply}

Janus started as a research prototype that we progressively refined to
industry-grade level; it is currently integrated in a commercial offer
targeting workplace safety. 

\fakeparagraph{A versatile firmware} Janus is designed as a
stand-alone, reusable firmware module, whose API sharply separates the
core functionality of reporting neighbors and their distance from the
application use.
Therefore, it can be exploited towards very different notions of
contact detection, e.g., supporting detection of crowds, and beyond
the context of social distancing, e.g., to enable proxemics studies or
proximity warning systems. 

Janus runs atop ContikiOS on the popular DWM1001C module by Decawave,
combining a Nordic nRF52832 SoC for MCU and BLE and a DW1000 UWB
radio. We place the latter in deep sleep mode whenever possible to
exploit its very low-power operation ($\sim$5nA), a task complicated
by the long delay ($\sim$5.5ms) to resume operation.

\fakeparagraph{A custom tag} We tested Janus on the Decawave MDEK1001
evaluation kits. These boards are equipped with USB ports and a nice
packaging, ideal for development and experimentation. Nevertheless,
their hardware is constrained; the integrated, energy-hungry Segger
debugger cannot be easily disabled, and LEDs provide the only form of user
feedback. These aspects, along with considerations about user comfort
when wearing the tag for prolonged periods, motivated the design of a
custom tag.

The current version has a badge form factor
($106 \times 64 \times 13mm$) and weighs $62g$. Inside the enclosure,
the hardware board includes the DWM1001C, a buzzer providing audible
and vibration user feedback, 2~LEDs, a multi-functional on/off
controller, and an $8\mathit{Mbit}$ Flash memory.  A rechargeable
$950\mathit{mAh}$ Lithium-Polymer battery powers the tag.

\fakeparagraph{A complete solution} In typical target domains like
large factories and offices, where tags enable both real-time alerting
and offline analysis, the core enabling functionality of Janus must be
supplemented by less innovative elements.

For instance, a \emph{gateway} enables data collection from the tags
via the UWB link and upload to the cloud, where data is persistently
stored and can be queried and visualized via a \emph{graphical
  dashboard}. From a hardware standpoint, the gateway is simply a
modified tag integrated with an embedded Linux-based system providing
Internet connectivity. The fixed gateways also provide \emph{coarse
  localization} near points of interest (e.g., a coffee machine), as
they can implicitly situate contacts in their neighborhood. A
\emph{crowd detection} feature is also built atop the Janus API,
raising an alarm when the number of neighbors is higher than a
configured threshold.

Finally, an effective and simple solution requiring no technical
knowledge is provided for situations where nodes are not used
continuously (e.g., only during work hours) and are amassed together
(e.g., at the concierge). Contact detection would be useless, wasting
energy; however, nodes detecting a special \emph{inhibitor} node
automatically enter a stand-by state for a predefined time (e.g.,
$5'$), after which only BLE is activated, to scan again; normal
operation is resumed only when the inhibitor is no longer found.

\section{What About Energy Consumption?}

Janus is designed with energy efficiency in mind. Battery replacement
or frequent recharging is a burden for personal use but becomes
unacceptable in companies, where hundreds or thousands of devices
carried by employees must be managed.

We investigated the lifetime of Janus by acquiring current
measurements with a Keithley SourceMeter~2450. Real-world scenarios
are a mix of periods where the user is alone and others where is in
contact; however, the exact proportions of the mix are obviously
unknown. To represent this, we use three measurement scenarios: when
a tag is alone and when in contact with exactly 1~and~9~others. The
first scenario serves as an upper bound for lifetime and as a building
block for the other two, for which we investigate different
proportions of alone \vs in-contact times, spanning several
operational conditions at once.

Moreover, we also examined different configurations representative of
typical use cases. Real-time alerting requires a configuration
ensuring a short contact detection latency; we set it to $2s$ as in the
industrial in-field deployments reported later. In many situations,
however, alerting is unnecessary or even distracting, e.g., when
worn by children at school. In these cases, only the monitoring and
recording of contacts matters and, given that typical recommendations
focus on relatively long contacts (e.g., $15'$~within~$2m$), higher
latencies are applicable; we study the values of $15s$ and $30s$ used in
our other in-field experiences.

\begin{figure}[!t]
  \centering
  \includegraphics[width=\columnwidth]{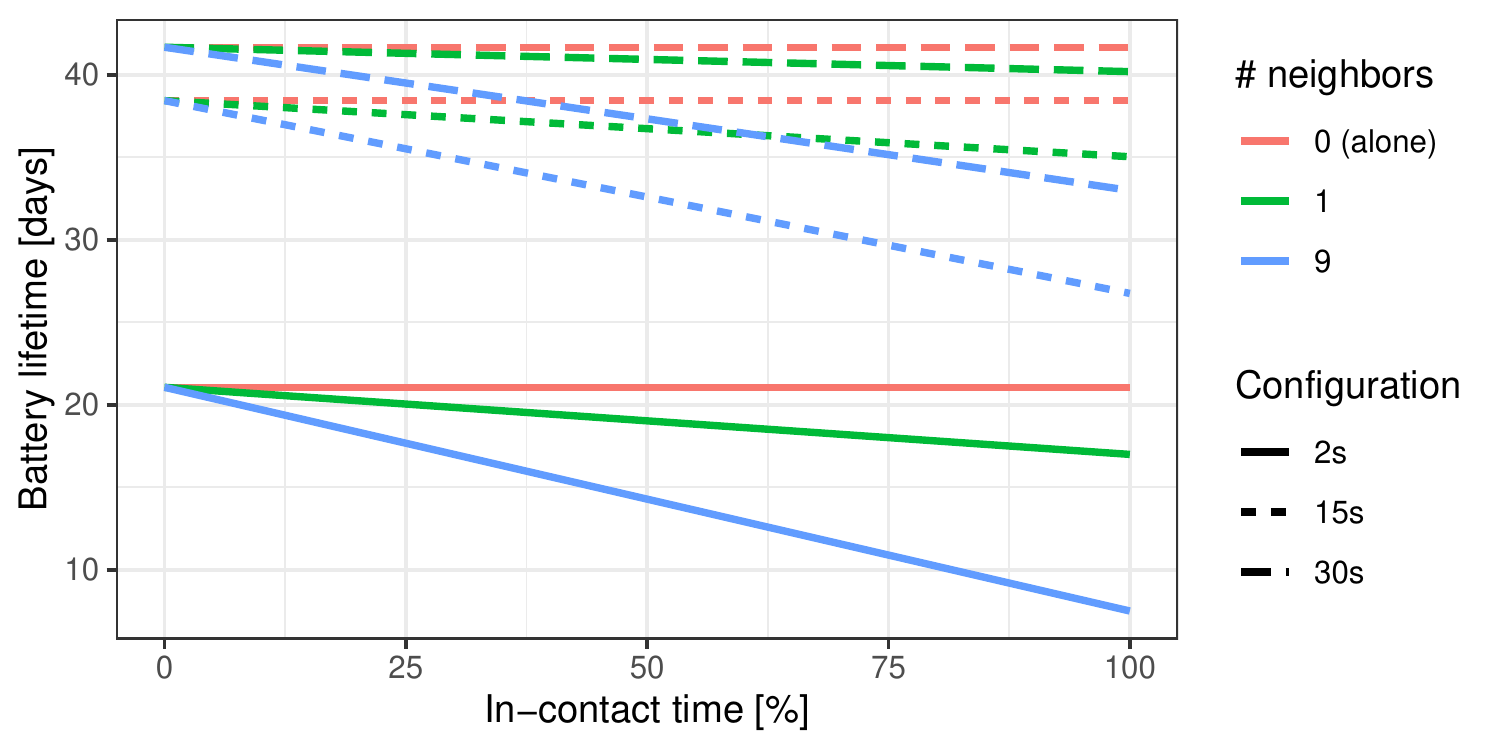}
  \caption{Estimated battery lifetime for a Janus tag \vs the percentage of time spent in that as a
    function of the time ratio spent in the communication range with
    one or nine other devices.}
  \label{fig:janusLifetime}
\end{figure}

Figure~\ref{fig:janusLifetime} shows the results, based on averages
over several 15-minute traces. When a tag is alone, only BLE is
active, scanning for neighbors; the average current draw ranges from
$1.88mA$ ($2s$) to $0.95mA$ ($30s$), yielding a lifetime from~21~to
41~days. When neighbors are present, the triggering of UWB increases
consumption, with a significantly different impact in the two use
cases. With a $30s$ latency, the current increases only to $0.985mA$
for 1~neighbor and $1.2mA$ for 9~neighbors; instead, the more reactive
configuration with $2s$ increases currents to $2.33mA$ and $5.28mA$,
respectively. These trends are reflected in the slopes of lifetime
curves (Figure~\ref{fig:janusLifetime}) that nonetheless confirm the
energy-efficiency of Janus; even with 9~neighbors \emph{continuously}
in contact, our tag lasts 7.5~days with $2s$ latency and 33~days with
$30s$. Note that this scenario is arguably an extreme one; in real
situations
\begin{inparaenum}[\em i)]
\item a user is rarely always in contact with a given number of users, and 
\item this number is usually much lower than~9, precisely due to the
  regulations about social distancing the tag is expected to support.
\end{inparaenum}
Therefore, in practice, the lifetime in each configuration is likely
somewhere between the 1- and 9-neighbor curves, and for an in-contact
time $<$100\%.

Interestingly, these values can be further improved, as the current
draw with both radios deactivated is relatively high, $0.72mA$. This
can be reduced by fine-tuning the interaction with peripherals and
other low-level aspects we did not address, as we focused on fully
optimizing the radio behavior. Still, even with this energy burden,
significant in relative terms, the lifetime reported is remarkably
higher than other research prototypes and market products.

Finally, these estimates assume 24-hour operation. When tags are worn
only during working hours and switched off otherwise, a significant lifetime 
increase can be achieved, e.g., threefold for an 8-hour workday.

\section{Janus in Action}
\label{sec:action}

We benchmarked Janus extensively against the requirements outlined
earlier in controlled, laboratory conditions; the same has been done
by independent evaluators in the context of a funded project. Results
confirmed the expected behavior \wrt accurate and timely contact
detection, and are omitted due to space constraints.

Here, we report on data gathered in several real-world contexts,
offering findings and insights about the practical application of
Janus.  Data was collected during the COVID-19 pandemic, with social
distancing and other safety measures in place. Proper procedures were
followed to recruit participants, compliant with GDPR and 
host organization regulations.

\begin{figure}
  \centering
  \includegraphics[width=\columnwidth]{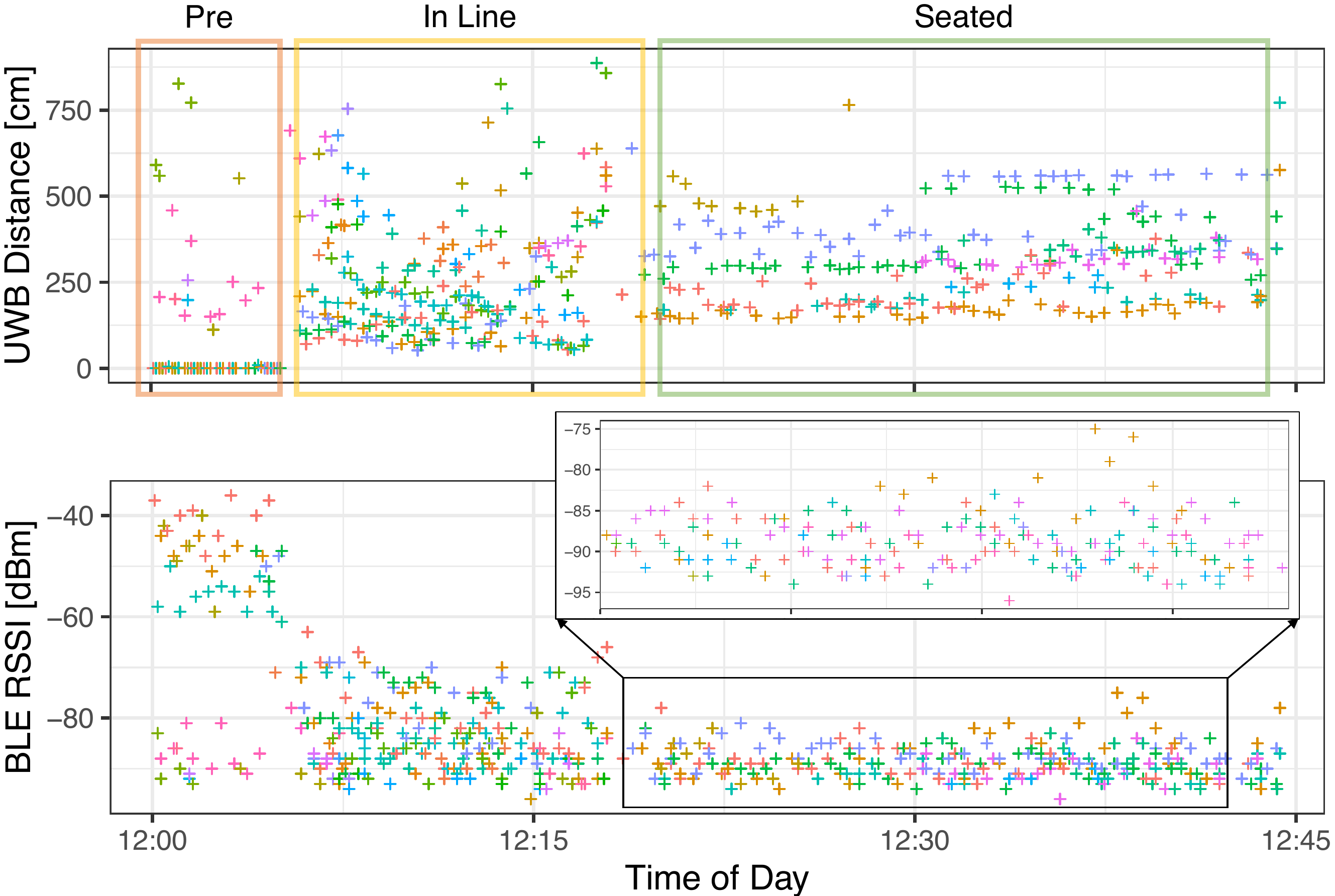}
  \caption{Cafeteria: Raw data from one individual. The zoomed-in area
    shows detail of BLE data.}
  \label{fig:mensa}
\end{figure}

\fakeparagraph{Cafeteria: Comparing BLE \vs UWB raw data} We begin
with a campaign in a company cafeteria where, over a 2-hour period, we
handed 90~workers a tag to carry during lunch. The dense setting is
challenging both to discovery and ranging; however, the inherent
flexibility of Janus allowed us to select an appropriate
configuration.  In the end, 148,768 samples $\langle$userID, RSSI,
distance, timestamp$\rangle$ were collected with a $30s$ latency, i.e.,
focusing on data collection rather than real-time
alerting. Figure~\ref{fig:mensa} shows the raw data of a single node;
each point denotes a measurement with a nearby device, itself
distinguished by color.

The UWB data (top) clearly shows three phases: when the node is ready
to be handed to the volunteer (\emph{Pre}), when the latter is waiting
to be served (\emph{\mbox{In Line}}), and when the volunteer is eating
(\emph{Seated}).  The distances between seated users are easily
discerned. This is not the case with BLE (bottom), even when zoomed in
to reveal detail.  Additional processing of RSSI values could improve
matters, as done by many BLE-based approaches; however, this
emphasizes that the raw data provided by UWB is \emph{immediately
  useful}.

\begin{figure}[!t]
  \centering
  \includegraphics[width=\columnwidth]{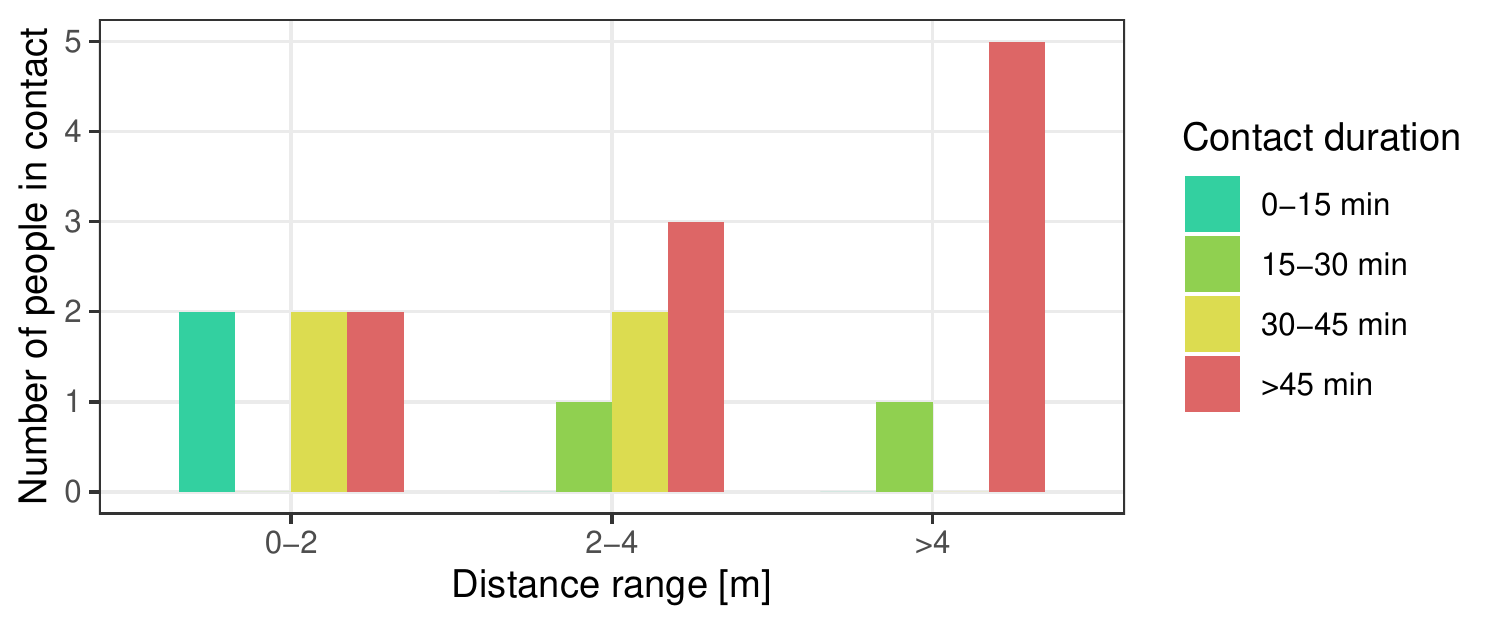}
  \vspace*{-7mm}
  \caption{Small office: Cumulative time of one individual at given
    distance from others during one workday.}
  \label{fig:group}
\end{figure}

\fakeparagraph{Same-office co-workers: Exploiting raw data} We report
data gathered with $15s$ latency from a typical office area where the
7~members of a research group are physically co-located.
Figure~\ref{fig:group} shows the \emph{cumulative} time one member
spent near others during one day, and highlights a potential
problematic situation: a significant amount of time ($>45'$) was spent
very near ($<2m$) two other members, and only slightly less ($30'$--$45'$)
very near two others. These times are derived straight from raw data,
by simply summing the $15s$ periods where a detection occurred. As
such, they do \emph{not} necessarily represent a (dangerous)
\emph{continuous} contact, whose definition we explore
next. Nevertheless, this further emphasizes that the accurate raw data
provided by Janus already offers actionable insights.

\fakeparagraph{Company-wide: Using a higher-level contact definition}
We now show results from an aggregation of the raw Janus data into a
higher-level notion of \textit{continuous contact}, often used to
characterize the risk of infection. We use the common definition of
risky contact as one occurring for at least $15'$ between individuals
within $2m$. We process raw data sequentially, looking at all distance
measurements between two individuals, regardless of direction. We
\emph{open} a contact when we first find a value within threshold, plus
a small tolerance ($20cm$) accounting for measurement
inaccuracies. We \emph{close} the contact when this condition becomes
continuously false for a given time period ($90s$); the last
value within threshold remains part of the contact.
The overall duration and average distance of the contact is then
computed, enabling a classification of contacts into:
\begin{itemize}
\item \emph{High} risk: below $2m$ for $>15'$;
\item \emph{Medium} risk: $<4m$ for $5'$ to $15'$ or between $2m$
  and~$4m$ for $>15'$;
\item \emph{Low} risk: otherwise. 
\end{itemize}
Although somewhat arbitrary, this classification is a realistic
example of how contact data could help prioritize actions.

\begin{figure}[!t]
  \centering
  \includegraphics[width=\columnwidth]{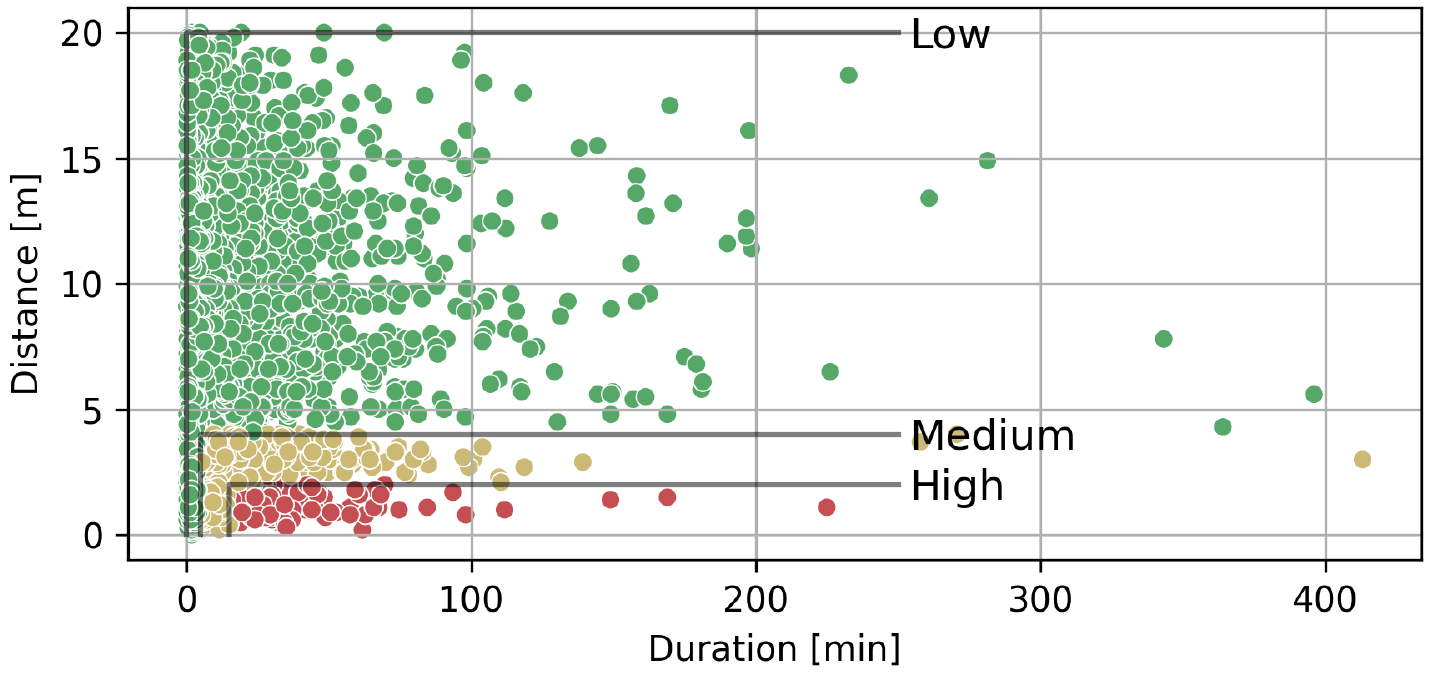}
  \vspace*{-7mm}
  \caption{Company-wide: Contacts of 90~individuals over 3~days.}
  \label{fig:fbkAll}
\end{figure}

To illustrate its power, enabled by Janus, we report 3~days of data at
$15s$ latency from 90~workers in a large company
building. Figure~\ref{fig:fbkAll} shows the duration \vs distance of contacts,
color-coded according to risk, providing a highly informative bird's-eye view.

Overall, a total of 5,899~minutes were recorded in high-risk contacts
over the 3~days. Although this seems large in absolute, on average it
is only $21.8'$ per person per day---about the same time users in the
above cafeteria scenario spent seated at lunch, potentially at risk.  Longer
accrued times were recorded at medium ($14,936'$) and low ($77,659'$) risk.

One can easily imagine follow-up analysis of this data, e.g.,
identifying the high-risk individuals, or analyzing the trends of
risky contacts throughout the day. Fixed nodes throughout the building
(e.g., at coffee machines) could also provide
approximate locations for some contacts.

\fakeparagraph{Factory floor: Real-time alerting and contact tracing}
We conclude by presenting data from 30~tags used on a factory
floor. The focus here was real-time alerting; tags are configured with
$2s$ latency. Further, data is gathered by a full-fledged product
(integrating Janus) at a customer site using
tags to record only high-risk contacts, offloaded via gateways
and available in cloud storage.

\begin{figure}[!t]
  \centering
  \includegraphics[width=\columnwidth]{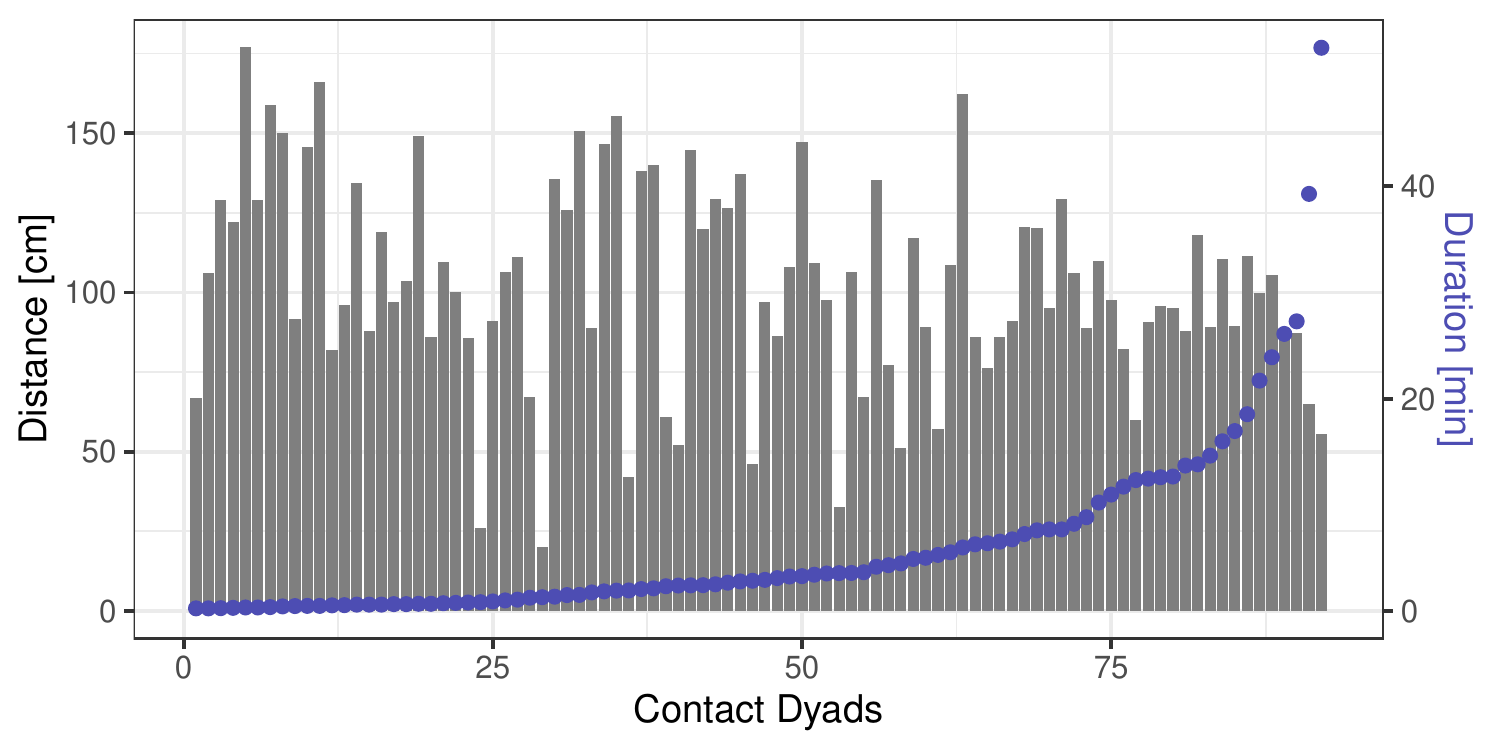}
  \vspace*{-7mm}
  \caption{Factory floor: average distance and total time in contact
    over a 24-hour period for each reported contact dyad.}
  \label{fig:replyData}
\end{figure}

We focus our attention on pairs (\emph{dyads}) of individuals, and
their total contact time in a day (Figure~\ref{fig:replyData}). If
tags $A$ and $B$ were within $2m$ for $6'$ in the morning and $9'$ in
the afternoon, the chart shows a point for dyad $A$--$B$ at $15'$,
with the corresponding histogram showing the average distance of the
dyad. For 30~individuals, there are 435~possible dyads; however, only
92 (21\%) were reported in contact. Of these, only 9~dyads exceed
$15'$ of total contact time. Further, these involve only 13~distinct
nodes, suggesting that long contacts are concentrated in few
individuals; this may be expected based on their duties, e.g.,
cooperatively moving heavy objects.

\fakeparagraph{Summary} The data we reported does not consider safety
measures mitigating risk, e.g., plexiglass dividers or masks;
accounting for them is an issue common to all contact
detection approaches, and outside the scope of this paper. Instead,
our data and analysis in various real-world scenarios confirm that the
energy-efficient protocol of Janus enables the reliable and flexible
collection of a wealth of accurate contact data, empowering
individuals and organizations with powerful and novel insights.

\section{Conclusions and Outlook}
\label{sec:ending}

We presented Janus, a novel dual-radio network protocol enabling
energy-efficient, timely, and accurate social contact detection among
devices equipped with both BLE and UWB.

Janus does not require an infrastructure and is highly and easily
configurable towards different application needs. These include
contact tracing analysis in the COVID-19 emergency, but are not
limited to it.  A prominent alternate use case are \emph{proximity
  warning systems} in industrial environments where workers must be
alerted of potential danger, e.g., operating machinery such as
forklifts and excavators in construction sites, or containers of
hazardous material.
In the context of social contact detection, Janus can also be
configured to interoperate with BLE-only approaches, e.g., GAEN-based
ones, enabling tags to record BLE advertisements from smartphones and
vice versa, with accurate ranging nonetheless limited to UWB-enabled
tags.

Nevertheless, the market penetration of UWB is rapidly increasing, as
witnessed by many smartphones from multiple vendors equipped with
it. As Janus does not rely on hardware-specific features of the radio
chips, we argue that the contribution described here is applicable to
existing and upcoming UWB devices, extending the applicability of our
solution to the wider user base and use cases enabled by smartphones.

\section*{Acknowledgment}

This work is partially supported by Fondazione VRT, by EIT Digital
(ProxyAware project, Activity 20666) and by the Italian government
(\mbox{NG-UWB} project, MIUR PRIN 2017).

\ifCLASSOPTIONcaptionsoff
  \newpage
\fi

\bibliographystyle{unsrt}

\end{document}